\documentclass[3p, onecolumn]{elsarticle}
\usepackage{graphicx} 
\usepackage{siunitx}
\usepackage{amsmath}
\usepackage{adjustbox}
\usepackage{hyperref}

\date{September 2024}

\title{Influence of Feedback Phase on Time Delay Signature and Chaos Bandwidth in a Laser subject to Dual Optical Feedback}
\author[add1]{Robbe de Mey\corref{cor1}}

\author[add1,add2]{Spencer W. Jolly}
\author[add1]{Martin Virte}
\affiliation[add1]{organization={Brussels Photonics, Department of Applied Physics and Photonics, Vrije Universiteit Brussel},
            addressline={Pleinlaan 2},
            postcode={1050},
            city={Brussels},
            country={Belgium}}
\affiliation[add2]{organization={Service OPERA-Photonique, Université libre de Bruxelles},
            city={Brussels},
            country={Belgium}}
\cortext[cor1]{Corresponding author: robbe.de.mey@vub.be}

\begin{document}
\begin{abstract}
Semiconductor lasers subject to optical feedback can behave chaotically, which can be used as a source of randomness. The optical feedback, provided by mirrors at a distance, determines the characteristics of the chaos and thus the quality of the randomness. However, this fixed distance also shows itself in the intensity, an unwanted feature called the Time Delay Signature (TDS). One promising solution to suppress the TDS is using double optical feedback. We study this system numerically in this paper. In particular, we focus on the impact of the feedback phase, a sub-wavelength change in the position of the mirrors, on the TDS and chaos bandwidth (CBW) of the system. We show that by precisely setting the feedback parameters, including the feedback phases, the TDS can be suppressed, and that the feedback phase control is necessary rather than optional to robustly suppress the TDS. In addition, it is possible to suppress the TDS without loss of the  CBW. At strong feedback rates the system can restabilize, and one can switch between a chaotic and steady state by changing only the feedback phase. Finally, we relate the feedback phase sensitivity to interference between the two delayed signals. This system is promising for applications of chaotic lasers as one can either suppress the TDS or increase the CBW.
\end{abstract}

\maketitle
\section{Introduction}
The output intensity of a semiconductor laser subject to optical feedback, simply by placing a mirror at a distance, can become chaotic~\cite{Ohtsubo2013, Uchida2012, Donati2013}. This is of interest for several applications, ranging from random number generation~\cite{Reidler2009}, safer communication~\cite{Udaltsov2005,Tronciu2008a}, or chaotic LiDAR~\cite{Chembo2019}. Underlying these applications is the high-bandwidth unpredictability of the intensity, inherent to a chaotic system. However, due to the fixed delay, a Time Delay Signature (TDS) appears in the intensity~\cite{Wei2022}, reducing the unpredictability which is essential for these applications. For example, the TDS shows itself in the time series hereby revealing information about the delay used in the optical feedback loop~\cite{Uchida2012}.

To suppress the TDS many promising solution have been proposed in literature~\cite{Wei2022}, including fiber-based solutions or optical injection. A fully integrated solution, on Photonic Integrated Circuits (PICs), would be highly beneficial for practical applications. As such, fiber-based solutions are not practical. Injection based solutions could work, but they are not that straightforward on every platform and, regardless, more complex. Moreover, optical isolators on PICs remain difficult to implement, and interplay between the master and follower laser could reduce the effectiveness of these methods. In this sense, a fully passive solution to suppress the TDS is desirable. 

One such passive way to reduce the TDS is by introducing a second optical feedback~\cite{Wu2009,Lee2005}. Adding a feedback loop can either stabilize or further destabilize the laser~\cite{Wu2009, Ruiz-Oliveras2006, Tavakoli2020, Tobbens2008, Liu1997, Sukow2002, Onea2019, Rogister1999, Barbosa2019}. By setting the second delay at a well-chosen delay close to the first delay it is possible to reduce the TDS~\cite{Wu2009}.  This result is of particular interest as they identified two cases where the TDS is strongly suppressed. For case I the two delays are close to each other and all signatures get suppressed. For case II one of the delays is at approximately half the other one, and only one of the signatures is suppressed. Using a second optical feedback loop might also be beneficial for increasing the CBW, for example in Ref.~\cite{Tronciu2008} it was shown that lower feedback strengths are needed to get high complexity chaos. 

In summary, using double optical feedback could increase the CBW and simultaneously suppress the TDS. However, in addition to recent literature, we have shown that the feedback phases have an impact on various aspects of the dynamics in this system~\cite{DeMey2022, DeMey2023}. We recently showed in an experiment that the feedback phase difference is an important parameter to control the TDS suppression~\cite{DeMey2022}. Here, we study this further to firmly confirm the importance of the feedback phase as a non-negligible parameter of the model. We numerically reproduce this observed behavior and further investigate the laser dynamics induced by double optical feedback, and discuss the interplay of all feedback parameters towards suppressing the TDS and increasing the CBW. In the second section, we introduce the system and discuss the relevant parameters and methods used. In the third section, we discuss the results. 

\begin{figure}[t]
	\centering
	\includegraphics[width = 0.5\linewidth]{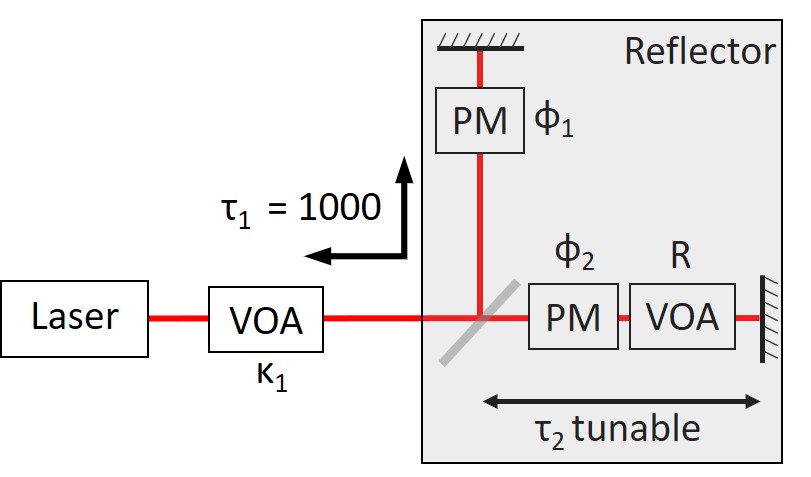}
	\caption{Schematic of the simulated system, a semiconductor laser coupled to two mirrors at a distance. The Variable Optical Attenuators (VOAs) control the amount of light fed back to the laser cavity. $\tau_1$ and $\tau_2$ determine the delay or equivalently the distance where each mirror is positioned. The control of the feedback phases, $\phi_1$ and $\phi_2$, is visualized as Phase Modulators (PMs).}
	\label{fig:figure1}
\end{figure}

\section{Simulation Model and Methods}
The simulated system is shown in Fig.~\ref{fig:figure1} where a semiconductor laser is coupled to two mirrors at a distance. To study this system we use the following normalized Lang-Kobayashi equations, extended for double optical feedback~\cite{Lang1980,Rogister1999}: 


\[ \dot{E}(t) = \frac{1}{2}(1-i\alpha)N(t)E(t)+ \kappa{}_{1} E(t-\tau_1)\exp(\phi_1) + \kappa{}_{2} E(t-\tau_2)\exp(\phi_2 ),\]

\[ \dot{N}(t) = (P-N(t)-(1+2N(t))E^2(t)))/T,\]

\noindent with $E$ the electric field, and $N$ the carrier density in the laser cavity. The rate equations are normalized with respect to the photon lifetime. Details for the variables used can be found in table~\ref{tab:LK}. For the laser parameters we use: $\alpha$ = 3, pump parameter $P = 1$ (corresponding to 2 times the laser threshold), and carrier lifetime $T = 1000$.

From the steady state solutions of the normalized equations without feedback ($\kappa_1 = \kappa_2 = 0$ ) we can calculate the Relaxation Oscillation Frequency (ROF) $v_{RO}$  of the solitary laser. Following the same approach as described in Ref.~\cite{Ohtsubo2013}, the ROF is: $v_{RO} = \frac{1}{2\pi}\sqrt{\frac{2P}{T}-\frac{1}{4}(\frac{2P+1}{T})^2}$. Taking the inverse and filling in $P = 1$ and $T = 1000$, we find the Relaxation Oscillation Period (ROP): $\tau_{RO} = 1/v_{RO} = 141$.

Each feedback arm has three parameters, the feedback rate ($\kappa_i$), the delay ($\tau_i$), and the feedback phase ($\phi_i$). We tune the feedback phase separately from the delay. This allows us to investigate the impact on the laser dynamics of large changes in the delay versus sub-wavelength changes. In an experimental setup, this could be implemented as a translation stage, capable of large ($\sim{}mm$) steps, to change the large-scale delay, and small ($\sim{}10$ $nm$) steps to change the delay on the sub-wavelength level, essentially changing the feedback phase. For example, as in the setup in our earlier work~\cite{DeMey2022}.

\begin{table}[t]
    \centering
        \resizebox{0.5\linewidth}{!}{\begin{tabular}{ l l l }
		Symbol & Represents & Typical value \\ \hline
		$\alpha$  & Linewidth enhancement factor & 3 \\  
		P  & Pump parameter & 1 \\
		T & Carrier lifetime & 1000    \\
		$\kappa_1$ & Feedback rate feedback loop 1 & 0--0.2\\
		$\tau_1$ & Delay feedback loop 1 & 1000 \\ 
		$\phi_1$ & Feedback phase feedback loop 1 & 0--2$\pi$ \\   
		$\kappa_2$ & Feedback rate feedback loop 2 & 0--0.2\\
		$\tau_2$ & Delay feedback loop 2 & 647--1500 \\ 
		$\phi_2$ & Feedback phase feedback loop 2 &  0--2$\pi$   \\  
		$R = \kappa_2/\kappa_1$ & Feedback rate ratio & /
	\end{tabular}}
	\caption{Details of the simulation parameters. In this paper we only change the feedback parameters ($\kappa_1$, $\tau_1$, $\phi_1$, $\kappa_2$, $\tau_2$, $\phi_2$). The laser parameters ($\alpha$, T, P) are fixed to these values.}
	\label{tab:LK}
\end{table}

The first feedback loop has a fixed delay of $\tau{}_{1} = 1000$, while the delay of the second feedback loop ($\tau_2$) is varied. The second feedback rate ($\kappa_2$) and the feedback ratio R are not independent parameters: $R = \kappa_2/\kappa_1$. As we focus on the influence of the feedback parameters on the laser dynamics, we will change the first feedback rate ($\kappa_1$), feedback phase 1 ($\phi{}_{1}$), the feedback rate ratio ($R$), delay 2 ($\tau{}_{2}$), or the feedback phase 2 ($\phi{}_{2}$). As a comparison, for a photon lifetime of 3 ps, a delay of $\tau_1 = 1000$ corresponds to a delay of $\tau'  = 3000$ ps, or a feedback length of approximately $l' = \tau'/2c \approx 45$ cm in free space.

\begin{figure*}[t]
	\centering
	\includegraphics[width = \linewidth]{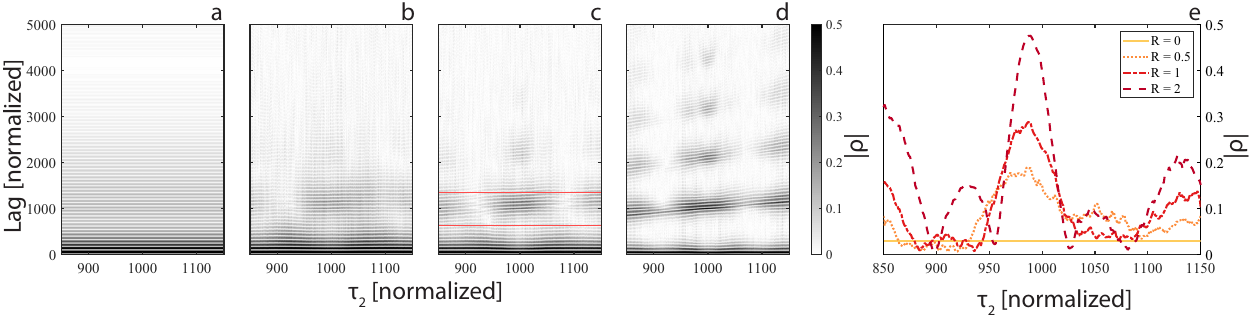}
	\caption{Evolution of the normalized ACF of the intensity time-series when sweeping $\tau_2$. (a)--(d) for increasing values of $\kappa_2$: R = 0, 0.5, 1, 2. The white zones indicate almost no correlation. (e) Intersection of (a)--(d) at lag = 1000. In (c) Red lines indicate the lag window we use for the TDS suppression. Simulation parameters: $\tau_1 = 1000$, $\kappa_1 = 0.01$, $\phi_1 = \phi_2 = 0$.}
	\label{fig:figure2}
\end{figure*}

To study the dynamics we analyze the time series of the amplitude squared of the electric field: $\vert E \vert ^2$, equivalent to the intensity. From this time series of intensity, we calculate the number of extrema. If there is only one extremum, the signal is constant and we consider the system to be in a steady state. If there are more extrema we consider the system is in a more complex state, from periodic to quasi-periodic and finally chaotic. In the case of 20 or more extrema we consider that the signal is chaotic, and this will be further motivated by the TDS and chaos bandwidth (CBW) analyses.

To identify the TDS, manifesting itself due to a fixed time delay in the system, we apply the normalized autocorrelation function (ACF) to the intensity time series. We calculate the ACF ($\rho$) as defined in Ref.~\cite{Wu2012}. As this function measures the linear correlation of the time series with a lagged version of itself, the TDS appears as a peak at the lag corresponding to the delay. Typically, we take the absolute value to focus on either the lack of correlation (i.e. values close to 0) or significant correlation (i.e. values close to 1). Although there exist other functions to measure the TDS~\cite{Wei2022}, for simplicity, we shall stick only to the ACF.

Besides suppressing the TDS we want to analyze the CBW of the system. We use this as an imperfect figure of merrit to study the impact of the feedback parameters on the chaos. It gives a rough but useful relative estimation of the unpredictability of the laser behavior~\cite{Lin2012}. To calculate it we take the RF spectrum from the intensity time series and only consider the bandwidth contributing to 80\,\% of the power, as described in Ref.~\cite{Lin2012}.

\section{Results and Discussion}

Ideally, in the context of chaos-based applications, the output of the laser should be chaotic with a non-detectable TDS while having the highest possible CBW. In the first subsection, we how the feedback phase comes into play when suppressing the TDS. In the second subsection we discuss the CBW. Finally, we study the stability of the system. We show that at high feedback rates, the system can go from stable to chaotic by a change of the feedback phase, implying that feedback phase is an essential parameter to describe the behavior of the system.

\subsection{Time Delay Signature Suppression}

An approach to suppress the TDS, without taking into account the feedback phases, was studied in Ref.~\cite{Wu2009}. The authors highlight that for case I, to suppress the TDS, the second delay should be close to the first delay. More specifically they mention that the suppression is strongest if $\Delta\tau = \tau_2-\tau_1 \approx -(3/2)\tau_\textrm{RO},-(1/2)\tau_\textrm{RO},(1/2)\tau_\textrm{RO}, (3/2)\tau_{\textrm{RO}}$, with $\tau_\textrm{RO}$ the ROP and for the case $R \approx 1$. They point out that the local minima in the ACF occur with a period of $\tau_\textrm{RO} / 2$, which is similar to the case of only one delay~\cite{Rontani2007}. This strategy can be thought of as using the second delay to damp the oscillations due to the first delay. However, the response of adding an additional delay is not necessarily linear, so this strategy should be carefully studied. We start from these results to have a baseline before taking all feedback parameters into account.

In the first simulation, we vary the second delay and confirm the impact on the TDS (Fig.~\ref{fig:figure2}). The feedback rate  is fixed, $\kappa_1 = 0.01$, the second feedback rate is increased and therefore $R$ increases. For all of these simulations, the number of extrema is more than or equal to 20, indicating a chaotic state. In Fig.~\ref{fig:figure2}(a)--(d) we plot the absolute ACF when varying $\tau_2$.  As we are increasing $R$ for these simulations but keeping $\kappa_1 = 0.01$, the total feedback rate is rising from Fig.~\ref{fig:figure2}(a)--(d). The first case (Fig.~\ref{fig:figure2}(a)) is a baseline, the second feedback rate is zero. In the third case, for $R = 1$, the feedback rates are equal. The case of $\tau_2 = 1000$ then is the same as the one delay case with twice the feedback rate. By increasing the second feedback rate the TDS can be suppressed, indicated by the regions of lower correlation in Fig.~\ref{fig:figure2}(b)--(d). Fig.~\ref{fig:figure2}(e) shows the ACF of (a)--(d) at lag = 1000.  Here, we compare the different plots for a line along lag = 1000, which shows that only at specific $\tau_2$ the TDS is more suppressed. The maximum height of the TDS depends on the total feedback rate, which was already shown in Ref.~\cite{Wu2009} and this is similar to the one delay case~\cite{Rontani2007}. Our simulations confirm that a second feedback can suppress the TDS~\cite{Wu2009}. In addition, even for different feedback ratios ($R$), the results remain qualitatively the same. 

However, focusing on only the ACF value at lag = 1000 is limiting, as even a signature close to the actual delay should be suppressed. Therefore, we will use a lag window in which we minimize the maximum ACF. We track in a window: $|\rho_\textrm{max}| = max(| autocorrelation(\vert E \vert ^2) |)$.  We take the total size of the window based on the ROP. As the period between minima is $\tau_{\textrm{RO}}/2$, we take the total size of the window equal to $5\tau_{\textrm{RO}}$ around lag =  $\tau_1$. For our specific system, this comes down to a window of $W = \tau_1 \pm 5\tau_{\textrm{RO}}/2 \approx 1000\pm 353$ (shown with red lines in Fig.~\ref{fig:figure2}(c)).

\begin{figure}[t]
	\centering
	\includegraphics[width = 0.5\linewidth]{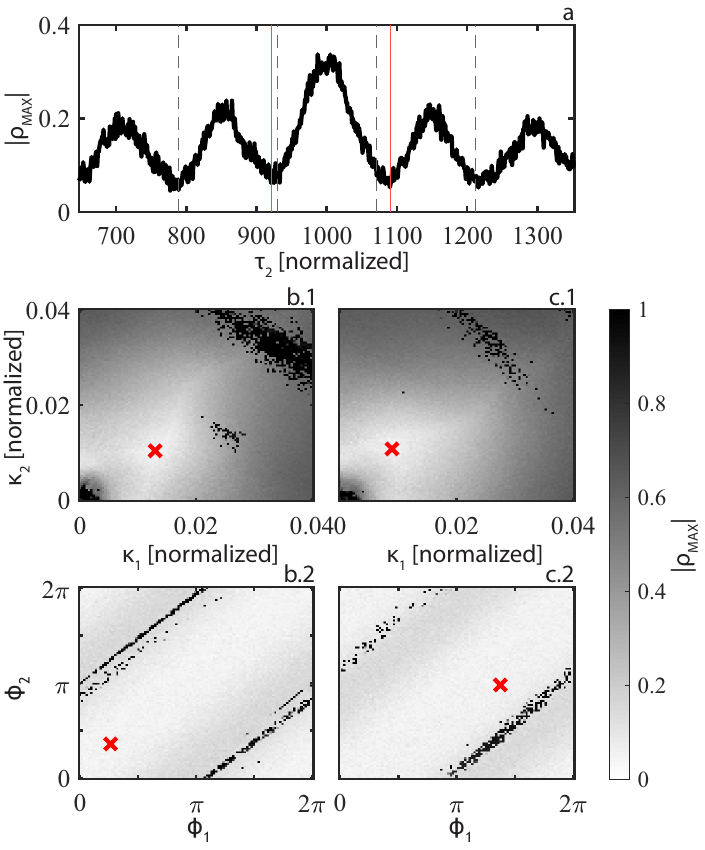}
	\caption{Suppressing the TDS by changing the feedback parameters. (a) Sweeping $\tau_2$, (b.1) and (c.1) sweeping the feedback rates, (b.2) and (c.2) sweeping the feedback phases. (a) dotted black lines indicate: $\Delta\tau = \tau_2-\tau_1 = -3\tau_\textrm{RO}/2,-\tau_\textrm{RO}/2,\tau_\textrm{RO}/2, 3\tau_{\textrm{RO}}/2$. Full red lines indicate the minimum value closest to $\tau_1$. In (b) $\tau_2 = 921$, (c) $\tau_2 = 1090$, where for (b.1) and (c.1) $\phi_1 = \phi_2 = 0$, and for (b.2) and (c.2) the feedback rates values used are indicated by the red crosses in (b.1) and (c.1). Red crosses are at the minimum of each map.}
	\label{fig:figure3}
\end{figure}

The results in Fig.~\ref{fig:figure2} show that the TDS depends on the second delay and feedback rate. On the other hand, we showed that the feedback phases are important for the stability~\cite{DeMey2023} and the TDS experimentally~\cite{DeMey2022}. To further study how to suppress the TDS with this system we take into account all feedback parameters ($\kappa_1$, $\tau_1$, $\phi_1$, $\kappa_2$, $\tau_2$, and $\phi_2$). To find a robustly suppressed TDS we apply the following steps:

\begin{enumerate}
	\item choose a second delay close to the first delay, around $|\Delta\tau| = |\tau_2 -\tau_1| < 5\tau_\textrm{RO}/2$, such that the TDS is maximally suppressed,
	\item sweep $\kappa_1$ and $\kappa_2$ to find a robust suppression of the main TDS peak,
	\item sweep $\phi_1$ and $\phi_2$ to find a robust suppression of the main TDS peak.
\end{enumerate}

\noindent In the first step, shown in Fig.~\ref{fig:figure3}(a), we want to find a TDS minimum to start from. We start from the case of $R = 1$ (Fig.~\ref{fig:figure2}(c)) and simulate for $\tau_2$ both smaller and larger than $\tau_1$. The dotted black lines indicate $\Delta\tau = \tau_2-\tau_1 = -3\tau_\textrm{RO}/2,-\tau_\textrm{RO}/2,\tau_\textrm{RO}/2, 3\tau_{\textrm{RO}}/2$, where we expect the optimal suppression based on Ref.~\cite{Wu2009}. This should be regarded as a rule of thumb as the lag at which the maximum TDS peak occurs shifts slightly depending on specific feedback parameters. Indeed, the red lines in Fig.~\ref{fig:figure3}(a), indicate that the local maximum suppression is at different values. In step 2, we sweep for a $\tau_2$ value below and above $\tau_1$: we take the local minima closest to $\tau_1$: $\tau_2 = 921$ and $\tau_2 = 1090$. For these points, the maximum ACF value in the lag window becomes respectively $|\rho_\textrm{max}| = \num{6.12e-2}$ and $|\rho_\textrm{max}| = \num{5.24e-2}$. The maximum TDS suppression when sweeping the second delay occurs for the minimum peak height around $\tau_2 \approx 800$ and $\tau \approx 1200$. However, the suppression there is only slightly better. In terms of hiding the TDS we assume having two delays closer together makes it more difficult to detect them both.

We analyse the TDS by sweeping both $\kappa_1$ and $\kappa_2$ and trying to find a more robust value for which a region around the main TDS peak is suppressed (shown in Fig.~\ref{fig:figure3}(b.1) and (c.1)). For the case of $\tau_2 = 921$, we find a minimal TDS at $\kappa_1 = \num{1.29e-2} $ and $\kappa_2 = \num{1.04e-2}$, with $|\rho_\textrm{max}| = \num{3.53e-2}$. For $\tau_2 = 1090$, we find the minimal TDS at $\kappa_1 = \num{8.89e-3}$ and $\kappa_2 = \num{1.08e-2}$, with $|\rho_\textrm{max}| = \num{4.72e-2}$. For both cases overall the ACF peak rises when increasing the feedback rates. In the black region in the top right of Fig.~\ref{fig:figure3} (b.1) and (c.1) the ACF peak takes on high values. For these regions, the number of extrema can be below 20, indicating that the system is not chaotic anymore. In addition, we note that the position of the strongest TDS peaks in the lag window latches on to the lag = $\tau_i$ of the strongest feedback loop, around $R = 1$ is the transition (supplementary Fig. 1). 

In the final step, we sweep the feedback phases, shown in Fig.~\ref{fig:figure3}(b.2) and (c.2). For the case of $\tau_2 = 921$ the minimum is reached for $\phi_1 = 0.825$ and $\phi_2 = 1.13$, for which $|\rho_\textrm{max}| = \num{2.76e-2}$. For the case of $\tau_2 = 1090$ the minimum is reached for $\phi_1 = 4.32$ and $\phi_2 = 3.08$, for which $|\rho_\textrm{max}| = \num{2.54e-2}$. By setting the feedback phases the TDS is further suppressed by a factor of two.

As indicated by the black regions, the maximum ACF peak can take on high values by just sweeping the feedback phases. This shows that to suppress the TDS with these delays, one needs to take into account the feedback phases. As similar ACF values mostly lie on diagonals, the main parameter that should be taken into account is the feedback phase difference. This confirms the experimental results in Ref.~\cite{DeMey2022}, where we show that the TDS peak in the ACF depends on the feedback phase difference. Of course, this effect works both ways: if, for some reason, an additional feedback phase term is added to the system, suppressing the TDS by choosing the second delay might only have a limited effect. Overall, the simulations indicate that this approach to suppress the TDS is quite robust. In the neighborhood of the optimal values (the red crosses in Fig.~\ref{fig:figure3} (b.1), (c.1), (b.2), and (c.2)) the TDS remains similarly suppressed.

\subsection{Chaos Bandwidth Increase}

\begin{figure}[t]
	\centering
	\includegraphics[width = 0.5\linewidth]{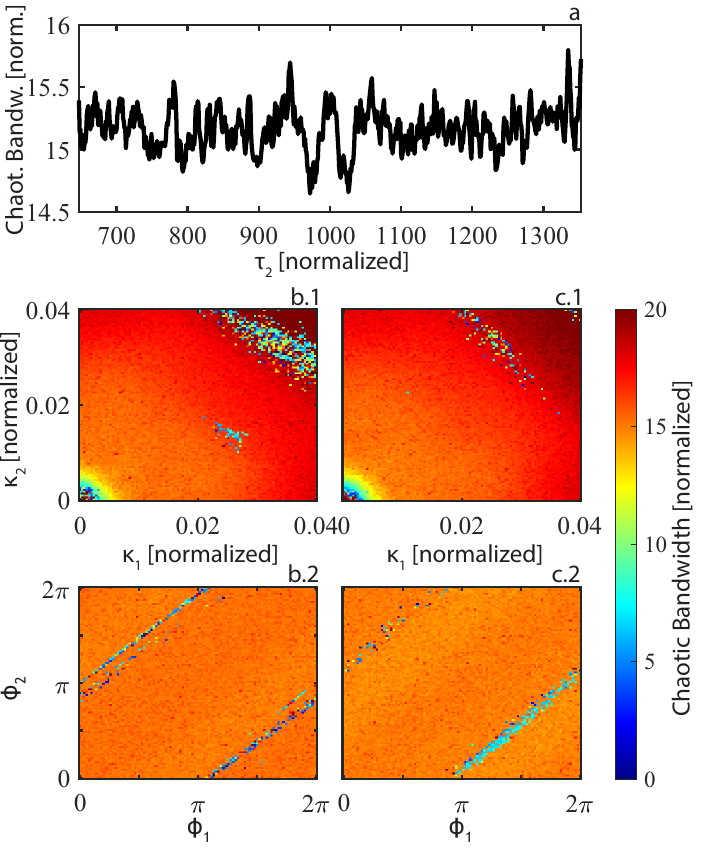}
	\caption{CBW when suppressing the TDS. (a) Varying $\tau_2$, (b.1) and (c.1) sweeping the feedback rates, (b.2) and (c.2) sweeping the feedback phases. The same simulation parameters as in Fig.~\ref{fig:figure3} are used.}
	\label{fig:figure4}
\end{figure}

\begin{figure}[t]
	\centering
	\includegraphics[width = 0.5\linewidth]{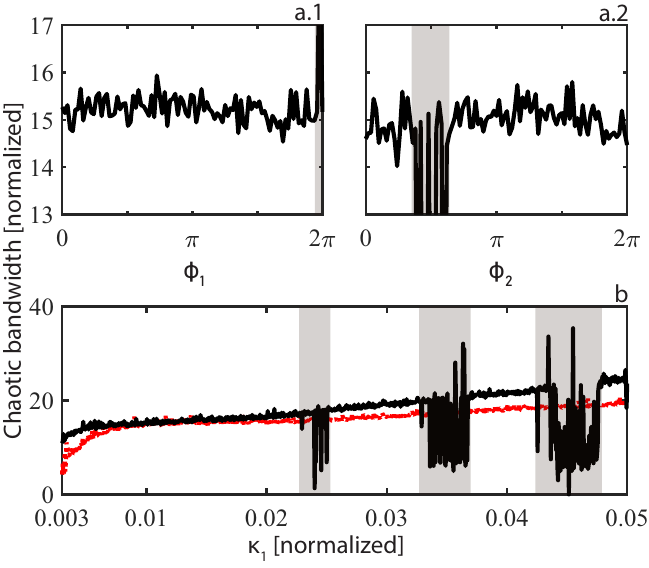}
	\caption{Evolution of the CBW versus the feedback parameters. (a)  Sweep of the feedback phases, around the optimum point of TDS in Fig.~\ref{fig:figure3} (c.2), (a.1) sweeping $\phi_1$, (a.2) sweeping $\phi_2$. (b) Sweep of the feedback rates. Black full line: sweeping feedback rates along optimal from Fig.~\ref{fig:figure3} (c.2), $R = 1.2$, $\tau_2 = 1090$, $\phi_1 = 4.32, \phi_2 = 3.08$. The dotted red line is the case for only one delay ($R = 0$, $\tau_1 = 1000$, $\phi_1 = 0$). In the gray areas the system can restabilize.}
	\label{fig:figure5}
\end{figure}

We study the same parameter regions as before but now focus on increasing the CBW (Fig.~\ref{fig:figure4}). When changing only $\tau_2$ (Fig.~\ref{fig:figure4}(a)) the total variation is below 10\,\% (data is smoothed with a moving average filter). Two local minima around $\tau_2 = 1000$ occur. When changing the feedback rate there is a clear impact on the CBW, shown in Fig.~\ref{fig:figure4}(b.1) and (c.1). Overall the CBW rises when increasing either $\kappa_1$ or $\kappa_2$. Increasing the feedback power increased the CBW. However, at high feedback rates, the same stability regions show themselves here. We shall discuss them further in the next subsection. When changing either feedback phase, shown in Fig.~\ref{fig:figure4}(b.2) and (c.2) there is a small impact on the CBW, although the quantitative change is small, on the order of 10\,\%. For the optimal TDS suppression case, the CBW is $15.31$ for $\tau_2 = 921$ and $15.19$ for $\tau_2 = 1090$. Changing the feedback parameters slightly around this optimum value has almost no impact on the CBW.

Fig.~\ref{fig:figure5}(a.1) and (a.2) show a line along the maps shown earlier. Here we only sweep one feedback phase while keeping the other one fixed at zero. There is a small impact due to the feedback phase in the order of 5\,\%, if the system does not restabilizes (indicated by large CBW jumps). However, the CBW increases significantly when either $\kappa_1$ or $\kappa_2$ is increased. Fig.~\ref{fig:figure5}(b) shows the case for $R = 0$ (dotted red line) and $R = 1$ (full black line). The CBW in this range for the case of two delays is always higher than the case of only one delay, although they are very close around $\kappa_1 = 0.01$. For the two delay case, the CBW rises faster with respect to the feedback rate. However, because $R = 1.2$, the energy returning to the laser rises faster, which could explain the discrepancy. Before $\kappa_1 = 0.003$, around $\kappa_1 = 0.025$, $\kappa_1 = 0.035$, and $\kappa_1 = 0.045$ (indicated by the gray areas) the system can have less than 20 extrema or the CBW takes on nonsense values. When discarding these regions the relation between the feedback rate and CBW is almost linear. We then apply a linear regression for each case from $\kappa_1 = 0.02$ to $0.05$. The slope of the one delay case is $\num{1.39e2}$ while the slope of the two delay case is $\num{2.62e2}$. The ratio between both is $1.88$, so the two delay case rises faster but not twice as fast. The case for $\tau_2 = 921$ is qualitatively the same.

In summary, our simulations show that the CBW mostly changes when changing the feedback rate, but is rather robust when either $\tau_2$ or the feedback phases are changing, as long as the system is still chaotic. However, even when changing the feedback rate the relative change in CBW is rather limited over the range where the TDS is most suppressed. On the other hand, to improve the CBW one can increase the feedback rate. Then the feedback phases should be adjusted to keep the system chaotic.

\subsection{Restabilization at high feedback rates}

To keep the system in a chaotic state, stable regions should be avoided. As shown above, for strong feedback rates the system can restabilize. Moreover, with a small change of the feedback phases the system can switch between chaotic and stable. In the previous section, the CBW increased with an increased feedback rate. If one wants to increase the CBW without interest in the TDS, increasing the feedback rate is an option. However, it then becomes essential to avoid the regions of stability. In this part, we investigate these regions of higher feedback rates to show when and why the restabilization occurs. 

\begin{figure}[t]
	\centering
	\includegraphics[width = 0.5\linewidth]{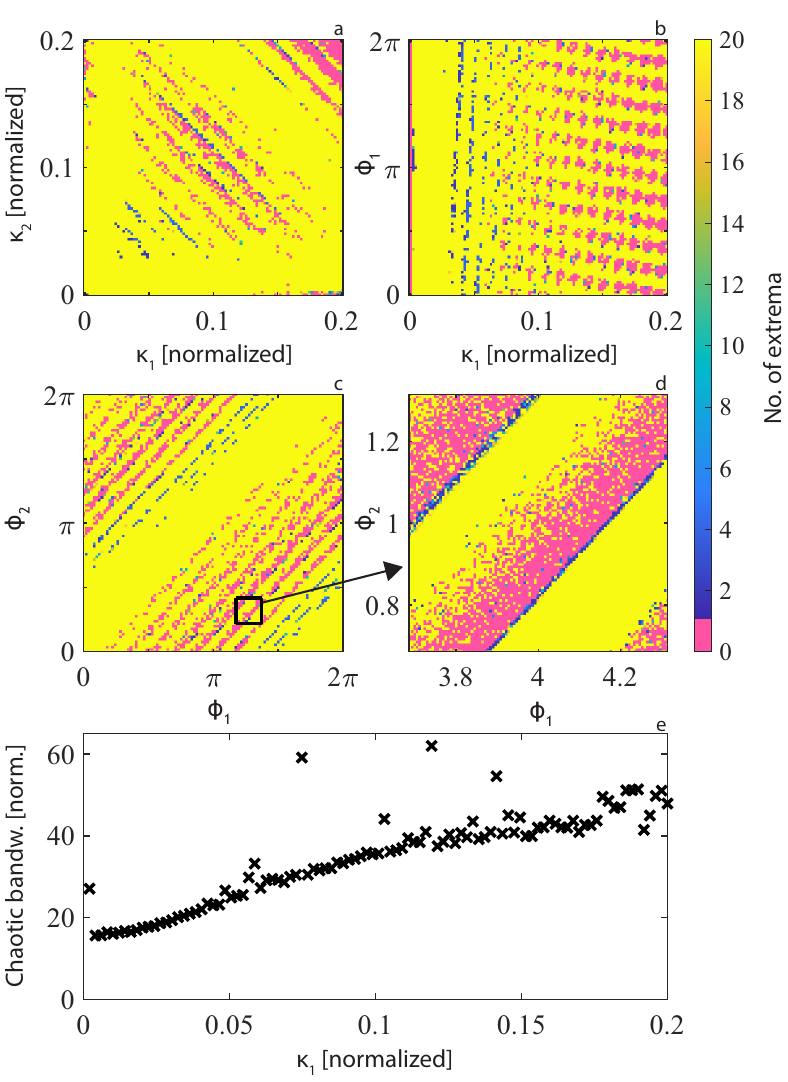}
	\caption{Restabilization at high feedback rates. Pink indicates a steady state, and yellow indicates a chaotic state. (a) Number of extrema when sweeping the feedback rates ($\kappa_1$ and $\kappa_2$). Simulation parameters:  $\tau_1 = 1000$, $\tau_2 = 1090$, $\phi_1 = \phi_2 = 0$. (b) Sweep along the diagonal ($R = 1$) of (a) and sweeping $\phi_1$ while keeping $\phi_2 = 0$. (c) Number of extrema when sweeping both feedback phases ($\phi_1$ and $\phi_2$). Simulation parameters:  $\tau_1 = 1000$, $\tau_2 = 1090$, $\kappa_1 = \kappa_2 = 0.1$. (d) Zoom in on the square in map (c). (e) The same parameters as in (b) are used but $\phi_1$ is chosen for the highest CBW.}
	\label{fig:figure6}
\end{figure}

First, we investigate the impact of the feedback rates on the stability. We start from a good position to suppress the TDS: $\tau_2 = 1090$. With the feedback phases set to zero, we then sweep the feedback rates to get the map shown in Fig.~\ref{fig:figure6}(a). Pink indicates a steady state (1 extremum) while yellow indicates a chaotic state ($\ge$ 20 extrema). For high feedback rates, the laser behavior can reach a steady state again. For these restabilizations to occur both $\kappa_1$ and $\kappa_2$ need to pass a certain threshold. After this threshold, when further increasing $\kappa_1$ and $\kappa_2$, the system can settle down to a steady state again.

These regions of restabilization are feedback phase dependent as shown in Fig.~\ref{fig:figure6}(b)--(d). Furthermore, above a certain feedback rate threshold, the laser system can always restabilize by setting the feedback phases. In Fig.~\ref{fig:figure6}(b) $\phi_1$ is swept while $\phi_2 = 0$ for each point along the $R = 1$ line in Fig.~\ref{fig:figure6}(a). By tuning the feedback phase difference the system can always become stable for $\kappa_1 \geq 0.059$. However, for an even lower threshold $\kappa_1 \geq 0.026$ the system settles down to a periodic state for specific feedback phase values. Above the first Hopf bifurcation and below this last threshold the system remains chaotic when sweeping the feedback phases, so to be used as a robust chaotic system you should remain in this region unless precise control of the mirror position is possible.  

Now, we fix the feedback rates ($\kappa_1 = \kappa_2 = 0.1$) and sweep both feedback phases. The results are shown in Fig.~\ref{fig:figure6}(c). By doing so, the laser switches between stable and chaotic. From our simulations, it seems to be a general feature that these restabilization regions, which are surrounded by chaotic regions in the feedback rates parameter space, are feedback phase dependent. We verified that similar behavior occurred in the presence of spontaneous emission noise. By zooming in on a part of the map, shown in Fig.~\ref{fig:figure6}(d), we see that there is a broad region of steady states. However, here and there the system can still become chaotic. The lower right part of this stability region has a small region where the laser behavior is periodic.

If the feedback phase can be controlled it is possible to reach very high CBW values. In Fig.~\ref{fig:figure6}(e) we show a sweep of $\kappa_1$ while choosing $\phi_1$ such that 1) the number of extrema is $\geq 20$ and 2) that the CBW is maximized (other parameters: $R = 1$, $\phi_2 = 0$). For feedback rate values up to $0.1$, we see an extension of the linear region discussed in the previous section. However, by setting the feedback phase we now avoid the restabilization regions. For higher feedback rates the CBW increases at a slower rate. For some specific feedback rates the chaos bandwidth takes on high values around 60, we consider these to be outliers. Understanding the nuances of this behavior is outside the scope of this paper.

\begin{figure}[t]
	\centering
	\includegraphics[width = 0.5\linewidth]{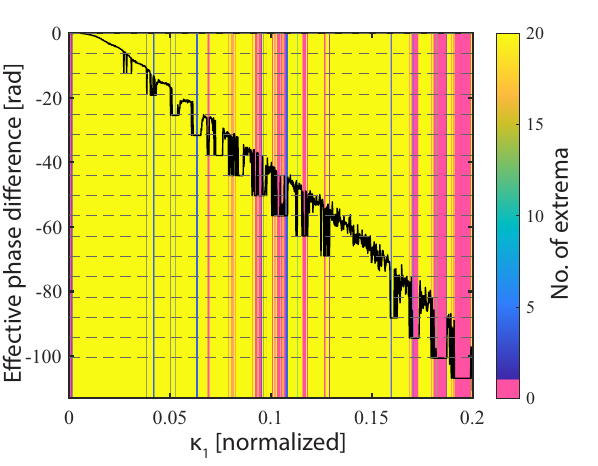}
	\caption{Evolution of the effective phase in the feedback section versus the stability of the system. Dotted straight lines indicate effective phase is in phase (multiple of $2\pi$). Simulation parameters: $\tau_2 = 1090$, $\kappa_2 = \kappa_1$, $\phi_1 = \phi_2 = 0$.}
	\label{fig:figure7}
\end{figure}

We hypothesize that these restabilizations occur due to constructive interference. Depending on the lasing wavelength and the length difference between the feedback arms, the light can interfere constructively. We support this by looking at the effective phase difference between both feedback sections when taking into account the shift in the laser wavelength as well as the delays and feedback phases, similarly defined in Ref.~\cite{DeMey2023}:

\begin{equation}
	\Delta\phi_\textrm{eff} = (\tau_2-\tau_1)\delta+\phi_2-\phi_1,
\end{equation}

\noindent in which we estimate the wavelength ($\delta$) by taking the average of $\delta = \frac{d\phi(t)}{dt}$. When looking at this effective phase difference (Fig.~\ref{fig:figure7}) the stable regions occur when both feedback loops are in phase, essentially when they are a multiple of $2\pi$. Here the system might act similar to a laser with one strong feedback loop. If the feedback phase is such that light from each feedback arm interferes constructively, the total feedback rate is high while the delay difference is small. If so, this case would be close to the case of the region V in the diagram of feedback regimes in a semiconductor laser~\cite{Donati2013, Tkach1986}. This should be confirmed with a more in-depth investigation.

In summary, to remain chaotic without sensitivity to feedback phase changes, you should remain at intermediate feedback rate values. Of course, if it's possible to control the feedback phases, a higher CBW can be reached by avoiding the stability regions.  We attribute this sensitivity to the feedback phases due to interference occurring in the feedback section. Although we only showed the simulations for one specific second delay value, this behavior occurs for a wide second delay range around the first delay.

\section{Conclusion}
We numerically studied a semiconductor laser coupled to two mirrors at a distance. We investigated the impact of the feedback parameters on the TDS and the CBW. The TDS can be suppressed when taking into account the feedback parameters. Building upon the work of Ref.~\cite{Wu2009} we confirm that when the delays are close to each other the TDS can be suppressed. However, we confirm here numerically that the feedback phases have to be controlled to effectively suppress the TDS, as we already showed for a particular case experimentally~\cite{DeMey2022}. Moreover, we show that control of the feedback phases is actually essential to suppress the TDS, of which a small shift can lead to a much higher TDS peak. By taking into account all the feedback parameters it is possible to strongly suppress the TDS.  

For the maximal suppressed TDS case we studied the CBW. We show that the TDS can be suppressed without loss of CBW. Changing the feedback parameters around the maximally suppressed TDS has almost no impact on the CBW. 

Within the studied parameter range, we find that our results show that the CBW increases significantly when increasing the feedback rates, as long as the feedback phase difference is controlled. In this context, we investigated the system for higher feedback rates. We show that above a certain threshold of the feedback rates, it is possible to switch from a chaotic to a steady state by changing the feedback phase. We use this to push the CBW much higher when controlling the feedback phase. We attribute this behavior to constructive interference occurring in the feedback section.

In conclusion, when taking into account all feedback parameters the CBW is high and the TDS suppressed. Or, if TDS suppression is less vital, the CBW can be increased. However, control of all feedback parameters is crucial, the feedback phase difference in particular. A system implemented on Photonic Integrated Circuits (PICs) could achieve this.

One parameter that should be investigated further in this context is the ROF of the laser. This parameter has an impact on the ease of TDS detection in the one delay case~\cite{Rontani2007} and comes into play in the dual delay system when setting the second delay. Further investigation in this context could lead to further TDS suppression and a higher CBW.

\section*{CRediT authorship contribution statement}
\textbf{Robbe de Mey:} Formal analysis (equal); Investigation (equal); Writing
– original draft (equal); Writing – review \& editing (equal).
\textbf{Spencer W. Jolly:} Formal analysis (equal); Supervision (supporting);
Validation (equal); Writing – review \& editing (equal). \textbf{Martin
Virte:} Formal analysis (equal); Methodology (equal); Supervision
(lead); Validation (equal); Writing – review \& editing (equal).

\section*{Declaration of competing interest}
The authors declare that they have no known competing financial interests or personal relationships that could have appeared to influence the work reported in this paper.

\section*{Data availability}
The data produced in this study have been deposited in the open access digital repository Zenodo. The data can be accessed at \url{https://doi.org//10.5281/zenodo.13833753}.

\section*{Acknowledgments}
This work was supported by the Fonds Wetenschappelijk Onderzoek (FWO) [project G0E7719N]; the Vlaamse Overheid, METHUSALEM program; the European Union H2020 research and innovation program, the Marie Sklodowska-Curie Action (MSCA) [project 801505].

\bibliographystyle{ieeetr}
\bibliography{biblo.bib}

\end{document}